\documentstyle[preprint,draft,floats,tighten,aps,prl,epsf,a4]{revtex}
\begin{document}
%
%
\preprint{\protect{
$\begin{array}{r}\text{IFUP-TH 2/94} \cr \text{HUTP-93/A038}\end{array}$}
}
\draft
\title{
The $b\rightarrow s\gamma$ decay revisited
}
\author{Giancarlo Cella\cite{auth::cella}, Giuseppe Curci\cite{auth::curci}}
\address{Dipartimento di Fisica dell'Universit\`a, Piazza Torricelli 2,
I-56126 Pisa, Italy, and Istituto Nazionale di Fisica Nucleare, Via Livornese
582/a I-56010 S.~Piero a Grado (Pisa), Italy}
\author{Giulia Ricciardi\cite{auth::ricciardi}}
\address{Lyman Laboratory of Physics, Harvard University, Cambridge, MA 02138}
\author{Andrea Vicer\`e\cite{auth::vicere}}
\address{Istituto Nazionale di Fisica Nucleare, Piazza Torricelli 2,
I-56100 Pisa, Italy}
\date{\today}
\maketitle
\begin{abstract}
In this work we compute the leading logarithmic corrections to the
$b\rightarrow s \gamma$ decay in a dimensional scheme which does not
require any definition of the $\gamma_5$ matrix. The scheme does not exhibit
unconsistencies and it is therefore a viable alternative to the t'Hooft
Veltman scheme, particularly in view of the next-to-leading computation. We
confirm the recent results of Ciuchini et al.
\end{abstract}
%
%
%
\section{Introduction}

The $b\rightarrow s \gamma$ decay has received 
considerable interest in the last years, thanks to its sensitivity to
physics beyond the Standard Model.

On the experimental side, the limit for the inclusive decay
based on the first observation of the $B\rightarrow K^\star \gamma$ 
transition by the CLEO collaboration~\cite{cleo:prl:71/93,McGill:Kim:CLNS93},
are compatible with the SM predictions, taking into account leading QCD
corrections.

On the theoretical side, after the first computations of the decay
amplitudes~\cite{grinstein:plb202,grigjanis:plb213,grigjanis:plb224,%
grinstein:npb339,grinstein:npb365}
and the confirmation of the results of~\cite{grinstein:plb202} given by us
in~\cite{us:plb248}, several improvements have occurred.

In~\cite{misiak:plb269,misiak:npb393} the analysis has been extended to
include all the relevant operators that mix under QCD corrections and
contribute to the leading logarithmic corrections.

In~\cite{ciuchini:plb316,ciuchini:rome93/973}
the long-standing problem of the differences between the results obtained
in~\cite{grinstein:plb202,us:plb248} and
in~\cite{grigjanis:plb213,grigjanis:plb224} has been solved, by
showing that the anomalous
dimension matrix is scheme dependent even at the leading order. This
effect is compensated by the matrix elements of $4$ fermion operators, and the
physical results are scheme independent. A similar analysis
limited to a subset of the relevant operators is also presented
in~\cite{misiak:TUM-T31-46/93}.

In~\cite{buras:TUM-T31-50/93} it has been performed an analysis of
all the sources of uncertainty in the leading logarithmic order computation (LLO)
and it has been shown that the inclusion of the next to leading order
corrections (NLO) can reduce the uncertainty from 25\% to less than 10\%.

In view of these considerations we think that it is time to undertake
a NLO computation and in this paper we reconsider the LLO
computation with a technique that we think particularly apt to this
purpose. As a byproduct, we confirm the recent results of Ciuchini et
al.~\cite{ciuchini:plb316,ciuchini:rome93/973}.

Dimensional methods differ in the
treatment of the $\gamma_5$ matrix: to our knowledge, the only method that
does not exhibit any inconsistency is the t'Hooft Veltman scheme
(HV)~\cite{tHooftVeltman}. Unfortunately, this method is very cumbersome and
in literature simplified versions have been used, like naive dimensional
regularization (NDR) or dimensional reduction (DRED)~\cite{siegel}. It has
been
shown~(\cite{buras:mpi-pae56-91,ciuchini:plb301,ciuchini:plb316,ciuchini:rome93/973}
and references therein) that NDR, DRED, HV methods give the same physical
results in a large class of radiative-correction computations (up to two
loops) where no trace ambiguities arise; on the other hand it is well known
that inherent pathologies of NDR and DRED schemes are present at three-loops
level~\cite{avdeev:chochia:vladimirov}.
We are therefore interested in a {\em practical} approach to the evaluation of
radiative corrections, as proposed in~\cite{curci:ricc:prd47}, avoiding the
extension of the $\gamma_5$ matrix to $d\neq 4$. This method appears to be
free of ambiguities, and in our opinion presents less complications than
the HV scheme, especially when performing computations by symbolic
manipulation programs.

The plan of the letter is as follows:
in section~(\ref{sec:strategy}) we describe the scheme
we have used, recalling the results in~\cite{curci:ricc:prd47},
then in section~(\ref{sec:bsgamma}) we show how we have applied the method to the
$b\rightarrow s \gamma$ process.
%
%
\section{Strategy}
\label{sec:strategy}
We aim to evaluate the amplitude for a decay process
induced by the exchange of virtual particles, such as the flavor changing
neutral current processes.

The integration of the heavy degrees of freedom (in our case, $W$ boson
and $t$ quark) leaves an effective hamiltonian, whose matrix elements in
the background of the ``light'' theory determine the 
decay amplitude
at the lowest
order in the heavy-mass expansion
\begin{equation}
\left<f\left|\right.i\right>_{\rm full} = \frac{1}{M^2}\,\sum_j
C_j\left(M,\,\mu\right) \left<f\left| O_j\right|i\right>_{\rm light}
\left(\mu\right) + O\left(\frac{1}{M^4}\right)\ .
\label{eq:lightAndHeavy}
\end{equation}
The coefficients of the operators in the effective hamiltonian
are determined matching the two sides of~(\ref{eq:lightAndHeavy}) 
in perturbation theory at a scale of the order of
the heavy masses. 
At $\mu\simeq M$ all the QCD large logarithms  originate
in the operator matrix elements of the 
effective theory 
 and the improved perturbation theory allows to resum them
sistematically.
The RG invariance of the l.h.s. of~(\ref{eq:lightAndHeavy}) is exploited,
together with the calculable $\mu$ dependence of the matrix elements, to
evolve down the r.h.s. at values of $\mu$ comparable with the scales of the
external states; this amounts to ``transfer'' the large logarithms from the
matrix elements to the coefficients.

The key element is therefore the evaluation of the anomalous dimension
matrix of the operators, which determines their $\mu$ evolution, together 
with the ``matching'' conditions at the $M$ scale and the matrix elements 
at the lower end of the evolution. This procedure is in general dependent on the
regularization and renormalization scheme chosen for the operators.

Now what is important for our discussion is the freedom to define the $d$
dimensional extension of the operators of the effective hamiltonian: as
in~\cite{curci:ricc:prd47} we will use this freedom to simplify the
computation.

Let us first observe that the general structure of the effective
hamiltonian for the $b\rightarrow s\gamma$ decay
contains chiral projectors $P_{L/R} = 1/2\left(1\mp\gamma_5\right)$;
therefore a definition of the $\gamma_5$ matrix is required in $d$
dimensions.
We have 
\begin{equation}
{\cal H}^1_{\rm eff} = \sum_i C^R_i R_i + \sum_i 
C^{LL}_i \left(L\otimes L\right)_i + \sum_i C^{LR}_i \left(L\otimes R\right)_i\ ,
\end{equation}
where $R_i$ stand for the magnetic momentum operators, while
$\left(L\otimes L\right)_i,\,\left(L\otimes R\right)_i$ are the
current-current operators: the field content is immaterial at this level.
Consider on the other hand the effective hamiltonian obtained with an
exchange $\gamma_5\rightarrow - \gamma_5$
\begin{equation}
{\cal H}^2_{\rm eff} = \sum_i C^R_i L^\prime_i + \sum_i C^{LL}_i
\left(R\otimes R\right)^\prime_i + \sum_i C^{LR}_i \left(R\otimes
L\right)^\prime_i
\end{equation}
where we intend that the operator $L^\prime_i$ is equal to $R_i$ apart
the chiral projector, and analogously for the other operators.

Since QCD does not know the sign of $\gamma_5$, the two hamiltonians will
receive the same perturbative corrections, that is
\begin{eqnarray}
{\cal D} N[L\otimes L_i] &+& \gamma^{LL\,R}_{i\,j} N[R_j] +
\gamma^{LL\,LL}_{i\,j} N[\left(L\otimes L\right)_j] +
\gamma^{LL\,LR}_{i\,j} N[\left(L\otimes R\right)_j] = 0\nonumber\\
{\cal D} N[\left(R\otimes R\right)^\prime_i] &+& \gamma^{LL\,R}_{i\,j}
N[L^\prime_j] + \gamma^{LL\,LL}_{i\,j} N[\left(R\otimes R\right)^\prime_j] +
\gamma^{LL\,LR}_{i\,j} N[\left(R\otimes L\right)^\prime_j] = 0\nonumber\\
&&\dots ,
\end{eqnarray}
assuming the same renormalization scheme is used.
The $N\left[{O}\right]$ symbol stands for the operator $O$ renormalized by 
minimal subtraction.
 
Let us now take the linear combinations which are symmetric and
anti-symmetric under the $\gamma_5$ sign flip
\begin{equation}
{\cal H}_{\text{s/a}} = \frac{1}{2}\left({\cal H}^1_{\rm eff} \pm {\cal H}^2_{\rm
eff}\right)\ .
\end{equation}
The original hamiltonian, which determines the physical amplitude, is a
combination of the symmetric and anti-symmetric hamiltonian; the key point
is that in order to obtain the scaling properties of ${\cal
H}^1_{\text{eff}}$ it is sufficient to know the QCD evolution of ${\cal
H}_{\text{s}}$.
In other words it is possible to choose the same renormalization scheme for
${\cal H}_{\text{s}}$ and ${\cal H}_{\text{a}}$, and therefore determine, for
instance, the evolution of an operator having chiral structure
$\left(V-A\right)\otimes\left(V\mp A\right)$ from the evolution of
$\left(V\otimes V\right) \pm \left(A\otimes A\right)$.
In the case at hand we can choose a scheme for the symmetric part
which is as simple as possible, and forget the anti-symmetric part.

Let us focus on the current-current operators; the extension of
the symmetrized hamiltonian ${\cal H}_{\text{s}}$ to $d$ dimensions can be
conveniently done reexpressing tensor products of $\gamma$ matrices as
elements of the Clifford $d$ dimensional algebra, for instance
\begin{equation}
\frac{1}{2}\left[\left(V-A\right) \otimes \left(V-A\right) +
\left(V+A\right) \otimes \left(V+A\right)\right]
\equiv \gamma_\mu \otimes \gamma_\mu + \frac{1}{3!} \gamma_{\mu\nu\rho} 
\otimes \gamma_{\mu\nu\rho}\ .
\label{eq:example}
\end{equation}
More generally a complete basis in $d$ dimensions is provided by the
operators
\begin{eqnarray}
\left(\gamma^{\left(n\right)} \otimes\gamma^{\left(n\right)}\right) &=&
\sum_{\mu_,\mu_2,\dots\mu_n} \gamma_{\mu_1,\mu_2,\dots\mu_n} \otimes
\gamma_{\mu_1,\mu_2,\dots\mu_n}\nonumber\\
\gamma_{\mu_1,\mu_2,\dots\mu_n} &=& \frac{1}{n!} \sum_{p\in \Pi_n} (-1)^p 
\gamma_{\mu_1} \gamma_{\mu_2}\dots\gamma_{\mu_n}\ .
\end{eqnarray}

This extension to $d$-dimensions does not require any definition of a $d$
dimensional $\gamma_5$ and therefore it avoids the complications due to the
splitting between $\varepsilon$-dimensional space and $4$-dimensional
space.
It also allows to treat relevant operators having $n\leq 4$ and
evanescent operators having $n \geq 5$ at the same way.

Many tecnical details will be explained in a more extended 
work~\cite{extended} where a detailed account of the computation
and the full set of formulas needed in addition to the ones 
in~\cite{curci:ricc:prd47} to implement the method will be given
as a reference for future works.
%
%
\section{The Leading Log corrections to the \protect{$b\rightarrow
s\gamma$} hamiltonian}
\label{sec:bsgamma}
Let us write down the effective on-shell hamiltonian for the $b\rightarrow s 
\gamma,\,b\rightarrow s g$, already used in ~\cite{misiak:plb269,ciuchini:plb316},
in euclidean notation:
\begin{eqnarray}
\label{eq:misiak-ciuchiniBasis}
{\cal H}_1 &=& \frac{G_F}{\sqrt{2}} V_{t\,s}^\star\,V_{t\,b}\sum_i C_i Q_i\nonumber\\
Q_1 &=& \left(\bar{s}_\alpha c_\beta\right)_{V-A}\left(\bar{c}_\beta
b_\alpha\right)_{V-A}\nonumber\\
Q_2 &=& \left(\bar{s}_\alpha c_\alpha\right)_{V-A}\left(\bar{c}_\beta
b_\beta\right)_{V-A}\nonumber\\
Q_{3,\,5} &=& \left(\bar{s}_\alpha
b_\alpha\right)_{V-A}\sum_{q=u,d,s,c,b}\left(\bar{q}_\beta
q_\beta\right)_{V\pm A}\nonumber\\
Q_{4,\,6} &=& \left(\bar{s}_\alpha
b_\beta\right)_{V-A}\sum_{q=u,d,s,c,b}\left(\bar{q}_\beta
q_\alpha\right)_{V\pm A}\nonumber\\
Q_7 &=& \frac{\left(-i e Q_d\right)}{\left(4\pi\right)^2} m_b \bar{s}
\sigma_{\mu\nu\:\left(V+A\right)}F_{\mu\nu} b \nonumber\\
Q_8 &=& \frac{\left(-i g_s\right)}{\left(4\pi\right)^2} m_b \bar{s}
\sigma_{\mu\nu\:\left(V+A\right)}\hat{G}_{\mu\nu} b\ .
\end{eqnarray}
Following the steps outlined in the preceding section, we consider the
on-shell basis, in $d$ dimensions
\begin{eqnarray}
\label{eq:basisD}
O_{\left(1,\,n\right)} &=&
\frac{1}{n!}\,\left(\bar{s}_\alpha\gamma^{\left(n\right)} c_\beta\right)\left(\bar{c}_\beta \gamma^{\left(n\right)} b_\alpha\right)\nonumber\\
O_{\left(2,\,n\right)} &=&
\frac{1}{n!}\,\left(\bar{s}_\alpha\gamma^{\left(n\right)} c_\alpha\right)\left(\bar{c}_\beta
\gamma^{\left(n\right)} b_\beta\right)\nonumber\\
O_{\left(3,\,n\right)} &=&
\frac{1}{n!}\,\left(\bar{s}_\alpha\gamma^{\left(n\right)} b_\alpha\right)\sum_{q=u,d,s,c,b} \left(\bar{q}_\beta \gamma^{\left(n\right)}q_\beta\right)\nonumber\\
O_{\left(4,\,n\right)} &=&
\frac{1}{n!}\,\left(\bar{s}_\alpha\gamma^{\left(n\right)} b_\beta\right)\sum_{q}
\left(\bar{q}_\beta \gamma^{\left(n\right)} q_\alpha\right)\nonumber\\
O_5 &=& \frac{\left(-i e Q_d\right)}{\left(4\pi\right)^2} m_b \bar{s}
\sigma_{\mu\nu}F_{\mu\nu} b \nonumber\\
O_6 &=& \frac{\left(-i g_s\right)}{\left(4\pi\right)^2} m_b \bar{s}
\sigma_{\mu\nu}\hat{G}_{\mu\nu} b\ .
\end{eqnarray}
The symmetrized operators corresponding to the ones
listed in~(\ref{eq:misiak-ciuchiniBasis}) are defined as
\begin{eqnarray}
\label{eq:on-shellBasis}
Q^s_1 &=& O_{(1,\,1)} + O_{(1,\,3)}\nonumber\\
Q^s_2 &=& O_{(2,\,1)} + O_{(2,\,3)}\nonumber\\
Q^s_{3/5} &=& O_{(3,\,1)} \pm O_{(3,\,3)}\nonumber\\
Q^s_{4/6} &=& O_{(4,\,1)} \pm O_{(4,\,3)}\nonumber\\
Q^s_7 &=& O_5\nonumber\\
Q^s_8 &=& O_6\ .
\end{eqnarray}

The first step is to determine the anomalous dimension matrix of the
operators listed in~(\ref{eq:basisD}), by using algebraic identities for
arbitrary $n$~\cite{avdeev,kennedy,curci:ricc:prd47}.

The next step is to determine the Renormalization Group evolution in $d$
dimensional space:
\begin{eqnarray}
\label{eq:rg}
{\cal D} N\left[{O_{\left(i,\,n\right)}}\right] &+& \sum_{j,\,m}
\gamma_{\left(i,\,n\right),\,\left(j,\,m\right)}
N\left[{O_{\left(j,\,m\right)}}\right]
+ \sum_{k} \gamma_{\left(i,\,n\right),\,k} N\left[{O_{k}}\right] = 0 \nonumber\\
{\cal D} N\left[{O_{n}}\right] &+& \sum_{m}\gamma_{n\,m} N\left[{O_{k}}\right] = 0
\end{eqnarray}
where 
\begin{eqnarray}
{\cal D} &=& \mu\frac{\partial}{\partial \mu} + \beta_{QCD}
\frac{\partial}{\partial g_S} + \gamma_{m_b}\, m_b \frac{\partial}{\partial
m_b}\nonumber\\
\beta_{QCD} &=& 2\,\varepsilon - g_s \frac{\alpha_s}{\left(4\pi\right)}
\,\frac{11\,C_A - 2\,n_f}{3}\\
\gamma_{m_b} &=& - \frac{\alpha_s}{\left(2\,\pi\right)}\,3\,C_F\ ;\nonumber
\end{eqnarray}
here and in the following $C_A,\,C_F$ are the Casimir invariants
of $SU(N_c)$, $C_A = 
N_c,\,C_F = \left(N_c^2 - 1\right)/\left(2\,N_c\right)$, and $n_f$ is the 
number of active flavors.
 
Note that the basis in~(\ref{eq:basisD}) is infinite, but the sums
in~(\ref{eq:rg}) run, at a definite order in the loop expansion, over a
limited set of operators.
For instance at one loop the sub-matrix $\gamma$ of operators $O_{1/2,n}$
results as follows, omitting a factor $\alpha_s / \left(4\,\pi\right)$
\begin{eqnarray}
&&\bordermatrix{
     &  (1,\,n-2) & (1,\,n) & (1,\,n+2) \cr
(1,\,n) & 
{{\left( 6 - n \right) \,\left( n - 5 \right) }\over {C_A}} & 
{\scriptstyle 4\,C_F\,\left( \left( 1 - n \right) \,\left( n - 3\right)\right)} &
{{-\left( \left( 1 + n \right) \,\left( 2 + n \right)  \right) }\over 
    {C_A}} \cr
(2,\,n) & 
{{\left( n - 6\right) \,\left( n - 5\right) }\over 2} &
{\scriptstyle 3\,\left( -2 + 4\,n - {n^2} \right)} &
{{\left( 1 + n \right) \,\left( 2 + n \right) }\over 2}\cr
 }\nonumber\\
&&\bordermatrix{
     &  (2,\,n-2) & (2,\,n) & (2,\,n+2) \cr
(1,\,n) &
\left( n - 6\right) \,\left( n - 5\right) &
0 &
\left( 1 + n \right) \,\left( 2 + n \right)\cr
(2,\,n) &
{{\left( 4\,C_F - C_A\right) \,\left( n - 6\right) \,
      \left(n - 5\right) }\over 2} &
{\scriptstyle \left(3\,C_A - 4\,C_F\right)\,\left( 2 - 4\,n + {n^2} \right)  - 
      4\,C_F} &
{{\left( 4\,C_F - C_A\right) \,\left( 1 + n \right) \,
      \left( 2 + n \right) }\over 2}
\cr
 }\ .
\end{eqnarray}
That means that even if the basis contains only operators
$O_{\left(x,\,1\right)},\,O_{\left(x,\,3\right)}$ at the $M_W$ scale, 
the evanescent operators $O_{\left(x,\,5\right)}$ appear at one-loop order.

The last step is to project the RG equations in $4$ dimensions.
This is most easily done by using the reduction formulas~\cite{bonneau} to
reexpress the insertion of evanescent operators in Green functions as a
contribution to the coefficients of relevant operators
\begin{equation}
N\left[{E_i}\right] = \sum_{i} r_{i,\,j} N\left[{R_j}\right]\ ,
\end{equation}
where the $\hat{r}$ matrix can be determined perturbatively. One is then 
able to decouple the evanescent operators at the level of the RG equation and 
this is equivalent, as one can easily check, to define a non-minimal subtraction 
scheme which sets to zero the matrix elements of evanescent operators,
as in~\cite{ciuchini:plb316}. 
The anomalous dimension matrix for relevant operators is modified as
follows (schematically)
\begin{equation}
\gamma_{r,\,r^\prime} \rightarrow \gamma_{r,\,r^\prime} + \sum_e
\gamma_{r,\,e} r_{e,\,r^\prime}\ . 
\end{equation}
At leading logarithmic order, only the reduction over magnetic momentum operators
is relevant, resulting in the following entries
\begin{equation}
\hat{r} = \frac{8}{15}\times \bordermatrix{
      & 5   & 6 \cr
(3,5) & 1   & 1 \cr
(4,5) & C_A & 0\cr
}\ ;
\end{equation}
the reduction over four fermion operators is of order $\alpha_s$ and it
is relevant only for the NLO computation.

We give the final results of the anomalous dimension matrix after the
reduction to $4$ dimensions in the basis~(\ref{eq:on-shellBasis}), in a
form suitable for comparison with Ciuchini et
al.~\cite{ciuchini:plb316,ciuchini:rome93/973}:
\begin{equation}
\hat{\gamma} = \frac{\alpha_s}{4\pi}\,\left(\begin{array}{cc}
\hat{\gamma}_{ff} & \hat{\gamma}_{fm} \cr
\hat{0} & \hat{\gamma}_{mm}
\end{array}\right)\ ,
\end{equation}
where $\hat{\gamma}_{ff},\,\hat{\gamma}_{mm}$ are the scheme independent anomalous
dimension matrices in the four-fermion and in the magnetic momentum sectors
respectively,
\begin{eqnarray}
\hat{\gamma}_{ff} &=& 
\bordermatrix{
  & 1 & 2 & 3 & 4 & 5 & 6 \cr
1 & -\frac{6}{C_A} & 6 & 0 & 0 & 0 & 0 \cr
2 &  6 & -\frac{6}{C_A} & -\frac{2}{3\,C_A} & \frac{2}{3} & -\frac{2}{3\,C_A}
& \frac{2}{3} \cr
3 & 0 & 0 & -\frac{22}{3\,C_A} & \frac{22}{3} & - \frac{4}{3\,C_A} &
\frac{4}{3} \cr
4 & 0 & 0 & \frac{2\left(9\,C_A - n_f\right)}{3\,C_A} & - \frac{2\left(9  - n_f\,
C_A\right)}{3\,C_A} & - \frac{2\,n_f}{3\,C_A} & \frac{2\,n_f}{3} \cr
5 & 0 & 0 & 0 & 0 & \frac{6}{C_A} & - 6 \cr
6 & 0 & 0 & - \frac{2\,n_f}{3\,C_A} & \frac{2\,n_f}{3} & - \frac{2\,n_f}{3\,C_A} &
\frac{2\left(n_f -18\,C_F\right)}{3}
}\ ,\nonumber\\
\hat{\gamma}_{mm} &=& 
\bordermatrix{
  & 7 & 8 \cr
7 & 8\,C_F & 0 \cr
8 & 8\,C_F & 16\,C_F - 4\,C_A
}\ ,
\end{eqnarray}
while $\hat{\gamma}_{fm}$ is the scheme dependent matrix connecting
dimension $6$ and dimension $5$ operators
\begin{equation}
\hat{\gamma}_{fm} =
\bordermatrix{
  & 7 & 8 \cr
1 & 0 & 6 \cr
2 & -\frac{232\,C_F}{9} & \frac{-8\,C_A}{3} + \frac{92\,C_F}{9} \cr
3 & -\frac{32}{3\,C_A} + \frac{32\,C_A}{3} + \frac{88\,C_F}{9} & 
-\frac{32}{3\,C_A} + \frac{32\,C_A}{3} + \frac{88\,C_F}{9} + 6\,n_f \cr
4 & \frac{64\,C_A\,C_F}{3} + \frac{4\,C_F}{9}\left(27\,\bar{n}_f -
4\,n_f\right) & -4 -\frac{8\,C_A}{3}\,n_f + \frac{92\,C_F}{9}\,n_f \cr
5 & \frac{32}{3\,C_A} -\frac{32\,C_A}{3} - \frac{40\,C_F}{3} &
\frac{32}{3\,C_A} - \frac{40\,C_F}{3} - 6\,n_f \cr
6 & -\frac{40\,C_A\,C_F}{3} - \frac{4\,C_F}{9}\left(4\,n_f +
27\,\bar{n}_f\right) & -8 +\frac{10\,C_A}{3} n_f - \frac{124\,C_F}{9} n_f
}\ ,
\label{eq:ADMfm}
\end{equation}
obtained from the computation of the graphs in Fig.~(1) for
different insertions of $4$ fermion operators.

\begin{figure}
\label{fig:fig1}
\begin{center}
\leavevmode
\epsfbox{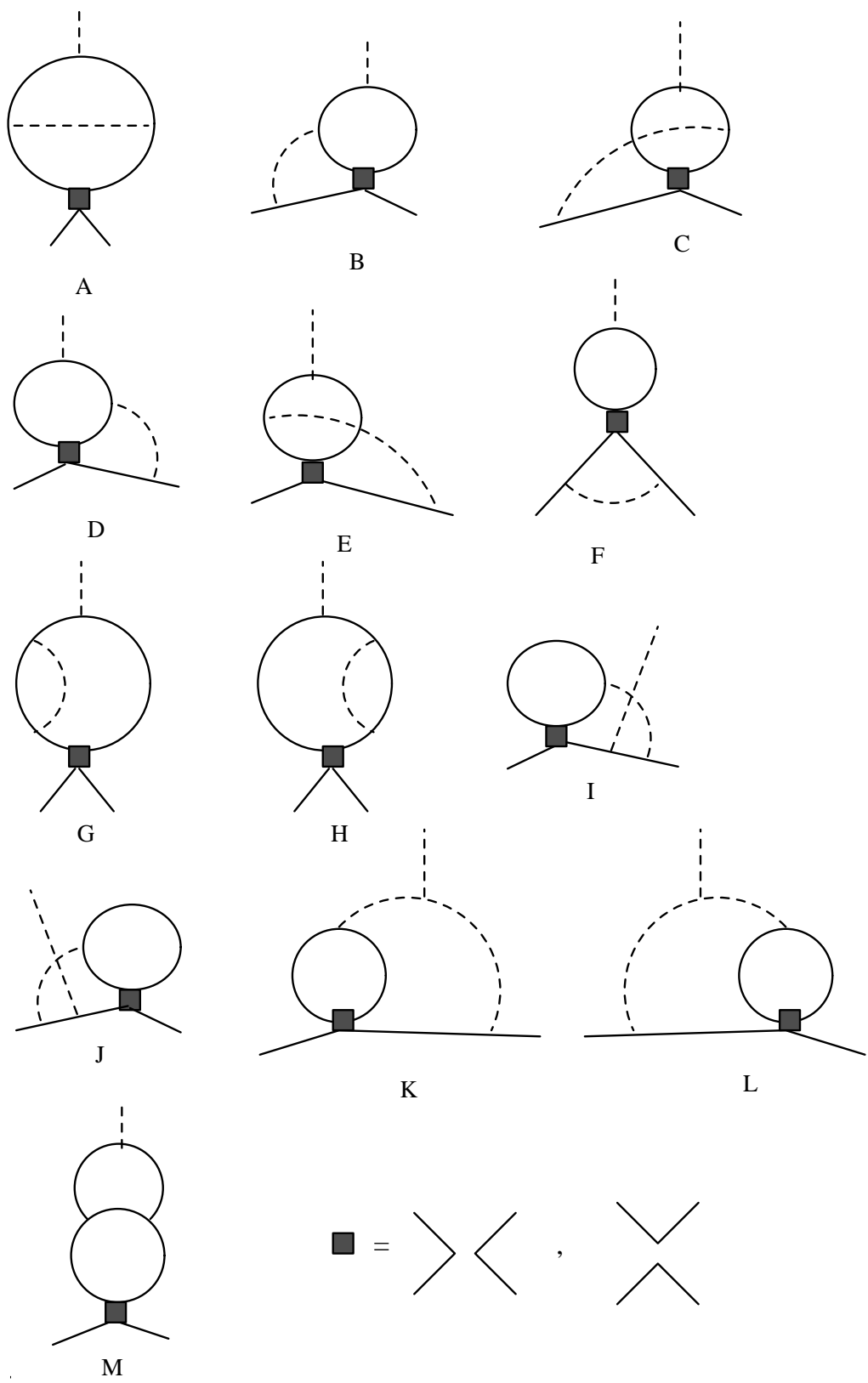}
\end{center}
\caption{Two loop graphs needed for $\hat{\gamma}_{fm}$}
\end{figure}

Following the authors of~\cite{ciuchini:plb316} we define $n_f = u +
d,\,\bar{n}_f = d - 2\,u$ and $u,\,d$ are the number of active up and down
flavors.

As shown in~\cite{ciuchini:plb316}, the matrix $\hat{\gamma}_{fm}$ is
scheme dependent, and in order to give a sensible result we have to compute
the matrix elements of operators $Q^s_{3,4,5,6}$:
\begin{eqnarray}
\label{eq:matel}
\left<s\,\gamma\left| \left(\begin{array}{c} Q^s_3\cr 
Q^s_4\end{array}\right)\right|b\right> &=& 
\frac{8}{3}\left(\begin{array}{c} 1\cr C_A\end{array}\right) 
\left<s\,\gamma\left|Q^s_7\right|b\right>\nonumber\\
 \left<s\,\gamma\left| \left(\begin{array}{c} Q^s_5\cr 
Q^s_6\end{array}\right)\right|b\right> &=& 
-\frac{2}{3}\left(\begin{array}{c} 1\cr C_A\end{array}\right) 
\left<s\,\gamma\left|Q^s_7\right|b\right>\nonumber\\
\left<s\,g\left| \left(\begin{array}{c} Q^s_3\cr 
Q^s_4\end{array}\right)\right|b\right> &=& 
\frac{8}{3}\left(\begin{array}{c} 1\cr 0\end{array}\right) 
\left<s\,g\left|Q^s_8\right|b\right>\nonumber\\
 \left<s\,g\left| \left(\begin{array}{c} Q^s_5\cr 
Q^s_6\end{array}\right)\right|b\right> &=& 
-\frac{2}{3}\left(\begin{array}{c} 1\cr 0\end{array}\right) 
\left<s\,g\left|Q^s_8\right|b\right>\ .
\end{eqnarray}
We have checked that the physical result is indeed scheme independent.
The easiest way to  show this in our context is to perform a finite
renormalization
\begin{equation}
N^\prime\left[{Q^s_i}\right] = \left(\hat{F}\right)_{i\,j}
N\left[{Q^s_j}\right]
\end{equation}
which sets to zero the matrix elements in~(\ref{eq:matel}). Note that 
this finite subtraction is independent on the coupling $\alpha_s$,
a consequence of the  fact that the scheme dependence
stems from penguin diagrams at zeroth  order in QCD~\cite{ciuchini:plb316}.

It is well known that the ADM matrix is modified as follows
\begin{equation}
\hat{\gamma}^\prime = \hat{F} \hat{\gamma} \hat{F}^{-1}\ .
\end{equation}
Using the results in~(\ref{eq:matel}) to define the matrix $\hat{F}$, 
the scheme dependent sub-matrix becomes:
\begin{equation}
\hat{\gamma}^\prime_{mf} =
\bordermatrix{
  & 7 & 8 \cr
1 & 0 & 6 \cr
2 & -\frac{208\, C_F}{9} & \frac{116\,C_F}{9} - 4\,C_A \cr
3 & \frac{232\,C_F}{9} & \frac{232\,C_F}{9} - 8\,C_A + 6 n_f \cr
4 & \frac{8\,C_F}{9}\,n_f + 12\,C_F\,\bar{n}_f & 12 +
\left(\frac{116\,C_F}{9} - 4\,C_A\right) n_f \cr
5 & -16\,C_F & 4\,C_A - 16\,C_F - 6\,n_f\cr
6 & \frac{8\,C_F}{9}\,n_f - 12\,C_F\,\bar{n}_f & -8 + \left(2\,C_A -
\frac{100\,C_F}{9}\right) n_f
}\ ,
\label{eq:ADMHV}
\end{equation}
which as expected coincides with the result in the HV
scheme~\cite{ciuchini:plb316}, where the matrix elements
in~(\ref{eq:matel}) are zero.
%
%
\section{Conclusions}
\label{sec:conclusions}
We have confirmed the results of Ciuchini et 
al.~\cite{ciuchini:plb316,ciuchini:rome93/973} by using a method which 
appears well suited for the NLO order computation.

We stress that our considerations are of practical nature: in the
evaluation of QCD corrections to electroweak processes
this technique appears less involved than the t'Hooft-Veltman scheme,
without introducing any potential ambiguity in the regularization.

\begin{figure}
\label{fig:fig2}
\begin{center}
\leavevmode
\epsffile{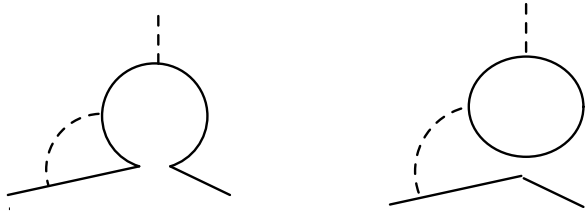}
\end{center}
\caption{Example of graphs related by generalized Fierz identities}
\end{figure}

Moreover this method presents other advantages, which will be elucidated in an 
extended description of the computation~\cite{extended}: we just 
mention here that, thanks to generalized Fierz identities~\cite{avdeev}, 
graphs like the ones in Fig.~(2) are related and
the knowledge of the ``bare'' graph with an open loop (on the left side of
Fig.~(2)) for arbitrary values of $n$ allows to determine directly the
value of the traced graph.

This and similar considerations  will be very useful in the NLO
computation, certainly a formidable task, but in our opinion worthwhile in
order to test as stringently as possible the Standard Model predictions.
\acknowledgments
One of us (G.~R) wants to thank Prof. H.~Georgi for interesting
discussions.
%
%
%

%
\end{document}